\documentclass[published]{nst}
\usepackage{subfigure,dcolumn}
\usepackage[T2A,T1]{fontenc}
\usepackage[russian,english]{babel}

\usepackage{listings}
\usepackage{epstopdf}
\begin{document}

\title{Production cross-sections of new superheavy elements with Z = 119-120 in fusion-evaporation reactions}\thanks{Supported by National Natural Science Foundation of China (No. 12105241, 12175072), Natural Science Foundation of Jiangsu Province (No. BK20210788), Jiangsu Provincial Double-Innovation Doctor Program (No. JSSCBS20211013) and University Science Research Project of Jiangsu Province (No. 21KJB140026) and Lv Yang Jin Feng (No. YZLYJFJH2021YXBS130) and the Key Laboratory of High Precision Nuclear Spectroscopy, Institute of Modern Physics, Chinese Academy of Sciences (No. IMPKFKT2021001)}

\author{Zi-Han Wang}
\affiliation{School of Physical Science and Technology, Yangzhou University, Yangzhou 225009, China}

\author{Peng-Hui Chen}
\email[Corresponding author,]{chenpenghui@yzu.edu.cn}
\affiliation{School of Physical Science and Technology, Yangzhou University, Yangzhou 225009, China}
\affiliation{Institute of Modern Physics, Chinese Academy of Sciences, Lanzhou 730000, China}

\author{Xiang-Hua Zeng}
\affiliation{School of Physical Science and Technology, Yangzhou University, Yangzhou 225009, China}
\affiliation{College of Electrical, Power and Energy Engineering, Yangzhou University, Yangzhou 225009, China }

\author{Zhao-Qing Feng}
\email[Corresponding author,]{fengzhq@scut.edu.cn}
\affiliation{School of Physics and Optoelectronics, South China University of Technology, Guangzhou 510641, China}

\begin{abstract}
We have calculated production cross sections of new superheavy elements with atomic number Z=119,120 in the fusion-evaporation reactions of $^{48}$Ca+$^{252}$Es, $^{48}$Ca+$^{257}$Fm, $^{49}$Sc+$^{252}$Es, $^{49}$Sc+$^{251}$Cf, $^{50}$Ti+$^{247}$Bk, $^{50}$Ti+$^{251}$Cf, $^{51}$V+$^{247}$Cm, $^{51}$V+$^{247}$Cf, $^{54}$Cr+$^{243}$Am, $^{54}$Cr+$^{247}$Cm, $^{56}$Mn+$^{244}$Pu, $^{56}$Mn+$^{243}$Am, $^{60}$Fe+$^{237}$Np, $^{60}$Fe+$^{244}$Pu, $^{61}$Co+$^{238}$U, $^{61}$Co+$^{237}$Np, $^{64}$Ni+$^{231}$Pa, $^{64}$Ni+$^{238}$U, $^{65}$Cu+$^{232}$Th, $^{65}$Cu+$^{231}$Pa, and $^{68}$Zn+$^{232}$Th within the dinuclear system model systematically.
The inner fusion barriers have been extracted from the driving potential or potential energy surface which could be used to predict the relative fusion probability roughly. 
The influence of mass asymmetry of the colliding partners on the production of new superheavy elements (SHE) has been investigated systematically. It is found that fusion probability increase along with the increasing mass asymmetry of colliding systems. The Ti-induced reactions have the largest cross-sections of the new SHE. The dependence of production cross-sections of new superheavy elements on the isospin of projectile nuclei has been discussed. The new SHE of $^{289-295}$119 has been predicted as the synthesis cross sections around serval picobarns in the $^{46,48,50,52}$Ti-induced reactions. Production cross-section of the element of $^{295}$120 has been evaluated as large as 1 picobarn in the reactions $^{46}$Ti ($^{251}$Cf, 2n) $^{295}$120 at $E^*$ = 26 MeV. The optimal projectile-target combinations and beam energies for producing new SHE with atomic number Z=119-120 are proposed for the forthcoming experiments.
\end{abstract}


\maketitle

\section{Introduction}

Synthesizing the new superheavy elements has obtained lots of attention from the theorists and experimentalists in the nuclear physics field since the " island of stability " of superheavy nuclei (SHN) has been proposed by the shell model in the 1960s\cite{RevModPhys.77.427}. The superheavy nuclei contain hundreds of nucleons and have been considered as a natural laboratory used to investigate the many body problems and structure of atomic under such a quite strong Coulomb force field. Currently, the experiments of synthesis of superheavy nuclei cost an amount of money and have a long experimental period, because of the extremely low cross section for synthesizing new superheavy elements. Therefore, it is necessary to make theoretical predictions for specific nuclei with the best projectile-target combinations and incident energy, formation cross-section, and evaporation channels, before performing such experiments.

On the experimental side, the fifteen superheavy elements (Z = 104-118) have been synthesized based on the hot-fusion and cold-fusion reactions at the laboratories in the past decades\cite{THOENNESSEN2013312}.  
Based on the hot-fusion reactions, the elements of rutherfordium, dubnium, seaborgium, flerovium, moscovium, livermorium, tennessine and oganesson have been synthesized firstly.
The element of rutherfordium (Rf, Z=104) was essentially discovered simultaneously in Dubna\cite{HEINLEIN1978407} and Berkeley\cite{GHIORSO197095} in the reactions $^{249}$Cf($^{12,13}$C, 3-4n)$^{257,258,259}$Rf.
The element of dubium (Db, Z=105) was essentially produced simultaneously in Dubna\cite{1970flog} and Berkeley\cite{PhysRevLett.24.1498} in $^{249}$Cf($^{15}$N,4n)$^{260}$Db and $^{243}$Am($^{22}$Ne, 4n)$^{261}$Db.
The element of seaborgium (Sg, Z=106) was essentially discovered at Berkeley\cite{PhysRevLett.33.1490} in $^{249}$Cf($^{18}$O, 4n)$^{263}$Sg.
The element of flerovium (Fl, Z=114) was discovered at Dubna\cite{PhysRevLett.105.182701} in $^{244}$Pu($^{48}$Ca, 3-6n)$^{286-289}$Fl. 
The element of moscovium (Mc, Z=115) was syntheiszed at Dubna\cite{PhysRevC.69.021601} $^{243}$Am($^{48}$Ca, 3n)$^{288}$Mc. 
The element of livermoriu (Lv, Z=116) was produced at Dubna\cite{PhysRevC.69.054607} in $^{245}$Cm($^{48}$Ca, xn)$^{293-x}$Fl.
The element of tennessine (Ts, Z=117) was discovered at Dubna\cite{PhysRevLett.104.142502} in $^{249}$Bk($^{48}$Ca, 3-4n)$^{293-294}$Fl. 
The element of oganesson (Og, Z=118) was syntheiszed at Dubna\cite{PhysRevC.74.044602} in $^{249}$Cf($^{48}$Ca, 3n)$^{294}$Og.
Based on the cold-fusion reactions, the elements of seaborgium, bohrium, hassium, meitnerium, darmstadtium, roentgenium, copernicium and nihonium have been synthesized firstly. 
The element of seaborgium (Sg, Z=106) was essentially discovered in Dubna\cite{1975Og} in the reactions of $^{207}$Pb($^{54}$Cr,2n)$^{259}$Sg.
The element of bohriumium (Bh, Z=107) was produced at GSI\cite{1981ZPhyA300107M} in $^{209}$Bi($^{54}$Cr, 1n)$^{262}$Bh. 
The element of hassium (Hs, Z=108) was essentially syntheiszed at GSI\cite{article84mu} in $^{208}$Pb($^{58}$Fe, 2n)$^{265}$Hs. 
The element of meitnerium (Mt, Z=109) was essentially discovered at GSI\cite{Mnzenberg1982} in $^{209}$Bi($^{58}$Fe, 1n)$^{266}$Mt. 
The element of darmstadtium (Ds, Z=110) was essentially produced at GSI\cite{articlehof95} in $^{208}$Pb($^{62}$Ni, 1n)$^{269}$Ds. 
The element of roentgenium (Rg, Z=111) was essentially discovered at GSI\cite{articlehof111} in $^{209}$Bi($^{64}$Ni, 1n)$^{272}$Rg. 
The element of copernium (Cn, Z=112) was essentially syntheiszed at GSI\cite{Hof96112} in $^{208}$Pb($^{70}$Zn, 1n)$^{277}$Rg. 
The element of nihonium (Nh, Z=113) was essentially produced at RIKEN\cite{doi:10.1143/JPSJ.73.2593} in $^{209}$Bi($^{70}$Zn, 1n)$^{278}$Nh. 
The earliest synthesis of superheavy elements with Z=104-118 are illustrated in Table \ref{tab1} as elements, reactions, evaporation channel, incident energy, isotope number, laboratory and year. In IMP Lanzhou, the superheavy isotopes of $^{258,259}$Db\cite{epjagan01}, $^{264,265,266}$Bh\cite{06-4-8npr06} and $^{271}$Ds\cite{cpl12zhang} have been synthesized. 

On the theoretical side, some sophisticated and practical theory models such as the time-dependent Hartree-Fock (TDHF) \cite{GUO2018401,10.3389/fphy.2019.00020,MARUHN20142195}, the improved quantum molecular dynamics (ImQMD) model\cite{PhysRevC.65.064608,PhysRevC.88.044611,PhysRevC.88.044605}, the dynamical approach based on Langevin equations\cite{PhysRevC.96.024618,PhysRevC.85.014608}, the dinuclear system model (DNS)\cite{FENG200650,PhysRevC.91.011603,PhysRevC.89.024615,epja20gga,07fengcpl} $etc.$ have been developed to describe the synthesis mechanisms of superheavy nuclei and predict their synthesis cross sections. 
The DNS model has been adapted which better involves the shell effect, dynamical deformation, fission, quasi-fission, deep-inelastic and odd-even effect. 
The DNS model has reproduced the excitation functions of the evaporation channels for the available experimental results in the Ref. \cite{PhysRevC.91.011603,FENG200650,FENG201082,07fengcpl}.

In this work, we have calculated all possible combinations of projectile-target for synthesizing the new superheavy elements Z=119,120 within the DNS model.
The aim of this paper is to investigate the influence of the entrance channel and projectile isospin on the production cross-sections of new superheavy elements in fusion-evaporation reactions and to predict the synthesis cross-section of new SHEs with Z=119,120. The article was organized as follows: In Sec. \ref{sec2} we give a brief description of the DNS model. Calculation results and discussions are presented in Sec. \ref{sec3}. The summary is concluded in Sec. \ref{sec4}.

\begin{table}[!ht]
\tabcolsep=.1 cm
\renewcommand\arraystretch{1.5}
\centering
\footnotesize
\caption{\label{tab1} The initial synthesized superheavy elements with atomic number Z=104-118 were illustrated as the produced method, reactions, cross sections, incident energy, existed isotope number, synthesized laboratories, year and reference. }
\begin{tabular}{ccccccccc}
\hline
  SHN & Reactions & C.S. & Energies  &  I.N. & Lab.  & Year   \\ \hline
Rf (104) & $^{12}$C + $^{249}$Cf   & xn\cite{HEINLEIN1978407} & 125 MeV &13 & Dubna & 1969    \\
   & $^{13}$C + $^{249}$Cf   & xn\cite{GHIORSO197095} & 135 MeV &   &  Berkeley  & 1969  \\ \hline 
Db (105) & $^{15}$N + $^{249}$Cf   &  4n
 \cite{PhysRevLett.24.1498} & 85 MeV &11 &  Berkeley & 1970 \\
   & $^{22}$Ne + $^{243}$Am   &  4n\cite{1970flog} & 114 MeV &   & Dubna   & 1970 
\\ \hline
Sg (106) & $^{54}$Cr + $^{207}$Pb   & 2n\cite{1975Og} & 262 MeV &12& Dubna  & 1974 \\
 & $^{18}$O + $^{249}$Cf   & 4n\cite{PhysRevLett.33.1490} & 95 MeV &  & Berkeley & 1974 
\\ \hline
Bh (107) & $^{54}$Cr + $^{209}$Bi   & 1n\cite{1981ZPhyA300107M} & 262 MeV &10 & GSI &  1981 
\\ \hline
Hs (108) & $^{58}$Fe + $^{208}$Pb   & 2n\cite{article84mu} & 291 MeV &12 & GSI &  1984 
\\ \hline
Mt (109) & $^{58}$Fe + $^{209}$Bi   & 1n\cite{Mnzenberg1982} & 299 MeV &7 & GSI &  1982 
\\ \hline
Ds (110) & $^{62}$Ni + $^{208}$Pb   & 1n\cite{articlehof95} & 311 MeV &8 & GSI & 1985 
\\ \hline
Rg (111) & $^{64}$Ni + $^{209}$Bi   & 1n\cite{articlehof111} & 318, 320 MeV &7 & GSI &  1995 
\\ \hline
Cn (112) & $^{70}$Zn + $^{208}$Pb   & 1n\cite{Hof96112} & 344 MeV &6 &GSI &  1996 
\\ \hline
Nh (113) & $^{70}$Zn + $^{209}$Bi   & 1n\cite{doi:10.1143/JPSJ.73.2593} & 352.6 MeV &6 & RIKEN &  2004 
\\ \hline
Fl (114) & $^{48}$Ca + $^{244}$Pu   & xn\cite{PhysRevLett.105.182701} & 352.6 MeV &5 & Dubna &  2004 \\ \hline
Mc (115) & $^{48}$Ca + $^{243}$Am   & 3n\cite{PhysRevC.69.021601} & 248, 253 MeV &4 & Dubna &  2004 
\\ \hline
Lv (116) & $^{48}$Ca + $^{245}$Cm   & xn\cite{PhysRevC.69.054607} & 243 MeV &4 & Dubna &  2004 
\\ \hline
Ts (117) & $^{48}$Ca + $^{249}$Bk   & xn\cite{PhysRevLett.104.142502} & 247, 252 MeV &2 & Dubna & 2011 
\\ \hline
Og (118) & $^{48}$Ca + $^{249}$Cf   & 3n\cite{PhysRevC.74.044602} & 251 MeV &1 & Dubna & 2006 
\\ \hline
\end{tabular}
\end{table}

\section{Model Description}\label{sec2}

The dinuclear system (DNS) model has been applied to describe low-energy reaction mechanism of heavy-ion collision as the molecular-like configuration of two colliding partners which keep their own individuality in the fusion process.
The DNS model has three components which are capture part, fusion part, survival part. For the capture part, the colliding partners overcome the Coulomb barrier to form the composite system. For the fusion part, the kinetic energy and angular momentum dissipate into the composite system to enable the nucleon transfer between the touching colliding partners.  All of the nucleons have been transferred from projectile nuclei to the target nuclei which could form the compound nuclei with a little excitation energy and angular momentum. For the survival part, the highly excited compound nuclei will be de-excited by evaporating the light particles. Based on the DNS model, the evaporation residual cross sections of superheavy nuclei are written as
\begin{eqnarray}
\sigma _{\rm ER}\left ( E_{\rm c.m.} \right ) =\frac{\pi \hbar ^2}{2\mu E{\rm c.m.}} \sum_{J=0}^{J_{\rm max}}(2J+1)T(E_{\rm c.m.},J)\nonumber \\P_{\rm CN}(E_{\rm c.m.},J)W_{\rm sur}(E_{\rm c.m.},J). 
\end{eqnarray}
Here, the penetration probability $T(E_{\rm c.m.},J)$ is calculated by the empirical coupling channel model\cite{FENG200650}. The fusion probability $P_{\rm CN}(E_{\rm c.m.},J)$ is the formation probability of compound nuclei\cite{PhysRevC.80.057601,PhysRevC.76.044606}. The survival probability $W_{\rm sur}$ is the probability of the highly excited compound nuclei survived by evaporating light particles. The maximal angular momentum was set as $J_{\rm max}$ = 30 $\hbar$ \cite{J.Mod.Phys.E5191(1996)}.

\subsection{ Capture probability}

The capture cross-section is written as
\begin{eqnarray}
\sigma _{\rm cap}(E_{\rm c.m.})=\frac{\pi \hbar ^2}{2\mu E_{\rm c.m.}}\sum_{J}^{}(2J+1)T(E_{\rm c.m.},J). 
\end{eqnarray}
Here, the $T(E_{\rm c.m.,J})$ is penetration probability evaluated by the Hill-Wheeler formula \cite{PhysRev.89.1102} within the barrier distribution function.
\begin{eqnarray} \label{hwt}
 && T(E_{\rm c.m.},J)= \nonumber \int   f(B) \\&& \frac{1}{1+\exp\left \{ -\frac{2\pi }{\hbar \omega (J)}\left [ E_{\rm c.m.}-B-\frac{\hbar^2J(J+1)}{2\mu R\rm_B^2(J) }  \right ]  \right \} }dB. 
\end{eqnarray}
Here $\hbar \omega (J)$ is the width of the parabolic barrier at $R_{\rm B}(J)$. The normalization constant is $\int f(B)dB=1$. The barrier distribution function is the asymmetric Gaussian form\cite{FENG200650,PhysRevC.65.014607}
\begin{eqnarray}\label{gbd}
f(B)=
\left\{\begin{matrix}
\frac{1}{N}\exp [-(\frac{B-B_{\rm m}}{\Delta_{1} } )]\ \ B<B_{\rm m},\\
  & \\\frac{1}{N}\exp [-(\frac{B-B_{\rm m}}{\Delta_{2} } )]\ \ B>B_{\rm m},
  &
\end{matrix}\right.
\end{eqnarray}
Here $\bigtriangleup _2$ = (B$_{0}$-B$_{\rm s}$)/2, $\bigtriangleup _1 $=$\bigtriangleup _2$-2 MeV, B$_m$=(B$_0$+B$_{\rm s}$)/2, B$_0$ and B$_{\rm s}$ are the Coulomb barriers of the side-side collision and the saddle point barriers in dynamical deformations\cite{PhysRevC.65.014607}. The interaction potential of two colliding partners is written as 
\begin{eqnarray}
V(\{\alpha\})=V{\rm _C}(\{\alpha\})+V_{\rm N}(\{\alpha\}) + V_{\rm def}
\end{eqnarray}
with 
\begin{eqnarray}
V_{\rm def} = \frac{1}{2} C_1(\beta_1-\beta^0_1)^2+\frac{1} {2}C_2(\beta_2-\beta_2^0)^2). \nonumber
\end{eqnarray}

The 1 and 2 stand for the projectile and the target. The $R=R_1+R_2+s$ and s are the distance between the center, surface of the projectile-target. The $R_1$, $R_2$ are the radii of the projectile and target, respectively. The $\beta_{1(2)}^0$ are the static quadrupole deformation. The $\beta_{1(2)}$ are the adjustable quadrupole deformation. The $\{\alpha\}$ stand for $\{R,\beta_1,\beta_1,\beta_2,\theta_1,\theta_2\}$. We assume that the deformation energy were proportional to their mass\cite{PhysRevC.65.014607}, namely, $C_1\beta_1^2/C_2\beta_2^2=A_1/A_2$. So we get one deformation parameter $\beta=\beta_1+\beta_2$. The stiffness parameters $C_{\rm i}(\rm i=1,2)$ is calculated by the liquid-drop model\cite{MYERS19661}.
The nucleus-nucleus potential is calculated by the double-folding method \cite{PhysRevC.80.057601,PhysRevC.76.044606, J.Mod.Phys.E5191(1996)}.
The Coulomb potential was derived by Wong's formula, as follows \cite{PhysRevLett.31.766}.



\subsection{Fusion probability}

In the diffuse process, the fragments probability are evaluated by solving a set of master equations. The term of probability $P(Z_1,N_1,E_1,t)$ contain proton number, neutron number of $Z_1,$ and $N_1$, and the internal excitation energy of $E_1$. The master equation is written as \cite{PhysRevC.76.044606,FENG201082,FENG200933}
\begin{eqnarray}
&& \frac{d P(Z_1,N_1,E_1,t)}{d t} =  \nonumber \\ && \sum \limits_{Z'_1}W_{Z_1,N_1;Z'_1,N_1}(t) [d_{Z_1,N_1}P(Z'_1,N_1,E'_1,t) \nonumber \\ && - d_{Z'_1,N_1}P(Z_1,N_1,E_1,t)] + \nonumber \\ &&
 \sum \limits_{N'_1}W_{Z_1,N_1;Z_1,N'_1}(t)[d_{Z_1,N_1}P(Z_1,N'_1,E'_1,t) \nonumber \\ && - d_{Z_1,N'_1}P(Z_1,N_1,E_1,t)] - \nonumber \\
 &&[\Lambda ^{\rm qf}_{A_1,E_1,t}(\Theta) + \Lambda^{\rm fis}_{A_1,E_1,t}(\Theta)]P(Z_1,N_1,E_1,t).
\end{eqnarray}
The $W_{Z_1,N_1,Z'_1,N_1}$ ($W_{Z_1,N_1,Z_1,N'_1}$) was the mean transition probability from the channel ($Z_1,N_1,E_1$) to ($Z'_1,N_1,E'_1$) [or ($Z_1,N_1,E_1$) to ($Z_1,N'_1,E'_1$)]. The $d_{Z_1,N_1}$ are the microscopic dimension corresponding to the macroscopic state ($Z_1,N_1,E_1$). The sum contains all possible numbers of proton and neutron for the fragment $Z'_1$, $N'_1$ own. However, only one nucleon transfer at one time was supposed in the model with the relation $Z'_1$ = $Z_1$ $\pm$ 1, and $N'_1$ = $N_1$ $\pm$ 1. The excitation energy $E_1$ was the local excitation energy $\varepsilon^*_1$ in the fragment ($Z'_1$, $N'_1$)S \cite{PhysRevC.27.590}. The sticking time was evaluated by the deflection function \cite{LI1981107}. The motion of nucleons in interaction potential was governed by the single-particle Hamiltonian, 
\begin{eqnarray}
H(t) = H_0(t) + V(t)
\end{eqnarray}
where the total single particle energy and interaction potential were
\begin{eqnarray}
H_0(t) &&= \sum _K\sum_{\nu_K} \varepsilon_{\nu_K}(t)\alpha^+_{\nu_K}(t)\alpha_{\nu_K}(t) \\ 
 V(t) &&= \sum _{K,K^{'}} \sum_{\alpha_K,\beta_{K'}} u_{\alpha_K,\beta_{K'}}\alpha^+_{\alpha_K}(t)\alpha_{\beta_K}(t) \nonumber  \\ 
 &&= \sum_{K,K'}V_{K,K'}(t). 
\end{eqnarray}
 The quantities $\varepsilon_{\nu K}$ and $u_{\alpha_K,\beta_{K'}}$ represent the single-particle energies and the interaction matrix elements, which is defined as the centers of colliding nuclei assumed to be orthogonal in the overlapping region. Then the annihilation and creation operators were time-dependent. The single-particle matrix elements were parameterized as
\begin{eqnarray}
&& u_{\alpha_K,\beta_K'} =  U_{K,K'}(t)  \\ && \times \left\{ \exp \left[- \frac{1}{2}( \frac{\varepsilon_{\alpha_K}(t) - \varepsilon_{\beta_K}(t)}{\Delta_{K,K'}(t)})^2 \right] - \delta_{\alpha_K,\beta_{K'}} \right\} \nonumber
\end{eqnarray}
Here, The calculation of the $U_{\rm K,K'}(t)$ and $\delta_{\alpha_{\rm K},\beta_{\rm K'}}(t)$ have been described in Ref.\cite{PhysRevC.68.034601}.
The proton transition probability was microscopically derived by 
\begin{eqnarray}
\label{trw}
&& W_{Z_{1},N_{1};Z_{1}^{\prime},N_{1}} = \frac{\tau_{\rm mem}(Z_1,N_1,E_1;Z_1^{\prime},N_{1},E_{1}^{\prime})} {d_{Z_1,N_1} d_{Z_1 ^{\prime},N_1}\hbar^2} \nonumber \\
&&  \times \sum_{ii^{\prime}}|\langle  Z_{1}^{\prime},N_{1},E_{1}^{\prime},i^{\prime}|V|Z_{1},N_{1},E_{1},i \rangle|^{2}.
\end{eqnarray}
The neutron transition probability has a similar formula. The memory time and the interaction elements $V$ could be seen in Ref.\cite{PhysRevC.80.057601}. 

The evolution of the DNS along the distance $R$ may lead to quasi-fission. Probability of quasi-fission was calculated based on the one-dimensional Kramers equation as \cite{PhysRevC.68.034601,PhysRevC.27.2063}.
In the fusion process, the highly excited heavy fragments might lead to fission where the fission probability is calculated by the Kramers formula. 

In the relaxation process of the relative motion, the DNS will be excited by the dissipation of the relative kinetic energy and angular momentum. The excited composite system opens a valence space $\Delta \varepsilon_K$ in fragment $K$ ($K$ = 1, 2) which has a symmetrical distribution around the Fermi surface. The nucleons in the valence space were actively enabled to transfer. The averages on these quantities are performed in the valence space, 
\begin{eqnarray}
\Delta \varepsilon_K = \sqrt{\frac{4\varepsilon^*_K}{g_K}},\quad
\varepsilon^*_K =\varepsilon^*\frac{A_K}{A}, \quad
g_K = A_K /12,
\end{eqnarray}
where the $\varepsilon^*$ is the local excitation energy of the DNS, which provide the excitation energy for the mean transition probability. There are $N_K$ = $g_{\rm K}\Delta\varepsilon_{\rm K}$ valence states and $m_{\rm K}$ = $N_{\rm K/2}$ valence nucleons in the valence space $\Delta\varepsilon_K$, which give the dimension
\begin{eqnarray}
 d(m_1, m_2) = {N_1 \choose m_1} {N_2 \choose m_2}.
\end{eqnarray}
The local excitation energy is given by
\begin{eqnarray}
\varepsilon^* = E_x - (U_{\rm dr}(A_1,A_2) - U_{\rm dr}(A_{\rm P}, A_{\rm T}))
\end{eqnarray}
Where the $U_{\rm dr}(A_1, A_2)$ and $U_{\rm dr}(A_P, A_T)$ were the driving potentials of fragments $A_1$, $A_2$ and $A_P$, $A_T$, respectively. The excitation energy $E_{\rm x}$ of the composite system was converted from the relative kinetic energy dissipation\cite{PhysRevC.76.044606}. The potential energy surface (PES) of the DNS is written as
\begin{eqnarray}\label{pes}
&&U_{\rm dr}(A_1,A_2;J,\theta_1,\theta_2) =  B_1+B_2 - B_{CN}-V^{\rm CN}_{\rm rot}(J) \nonumber\\&&
+V_{\rm C}(A_1,A_2;\theta_1,\theta_2) + V_{\rm N}(A_1,A_2;\theta_1,\theta_2)
\end{eqnarray}

 Here $B_{i}$ ($i$ = 1, 2) and $B_{\rm CN}$ were the binding energies. The $V_{\rm rot}^{\rm CN}$ is the rotation energy of the compound nuclei. The $\beta_{\rm i}$ represent the quadrupole deformations of binary fragments. The $\theta_{\rm i}$ denotes collision orientations. The $V_{\rm C}$ and $V_{\rm N}$ are the Coulomb potential and nucleus-nucleus potential respectively.

By solving a set of master equations, the probability of all possible fragments is presented. The hindrance in the fusion process is named inner fusion barrier $ B_{\rm fus}$ which is defined as the difference from the injection position to the B.G. point. These fragments overcome the inner barrier which could fusion. Therefore, the fusion probability is evaluated by summing all of the fragments which could penetrate the inner fusion barrier. The fusion probability is evaluated by  
\begin{eqnarray}
P_{\rm CN}(E_{\rm c.m.},J,B)=\sum _{A=1}^{A _{\rm BG}} P(A,E_1,\tau_{\rm int}(E_{\rm c.m.},J,B)).
\end{eqnarray}
Here, the interaction time $\tau\rm_{int}(E_{c.m.},J,B)$ was obtained from the deflection function method \cite{PhysRevC.27.590}. 
We calculated the fusion probability as
\begin{eqnarray}
P_{\rm CN}(E_{\rm c.m.},J)=\int f(B)P_{\rm CN}(E_{\rm c.m.},J,B)dB
\end{eqnarray}
The Coulomb barrier distribution function $f(B)$ is taken as Eq. \ref{gbd}, so the fusion cross section is written as
\begin{eqnarray}
\sigma _{\rm fus}(E_{\rm c.m.})=\sigma _{\rm cap}(E_{\rm c.m.}) P_{\rm CN}(E_{\rm c.m.},J)  
\end{eqnarray}

\subsection{ Survival probability}

The compound nuclei formed by the all nucleons transfer from projectile nuclei to target nuclei which have a few excitation energies. The excited compound nuclei were extremely unstable which would be de-excited by evaporating $\gamma$-rays, neutrons, protons, $\alpha$ $etc.$) against fission. The survival probability of the channels x-th neutron, y-th proton and z-alpha was expressed as \cite{Chen_2016,FENG201082,FENG200933}
\begin{eqnarray}
&&W_{\rm sur}(E_{\rm CN}^*,x,y,z,J)=P(E_{\rm CN}^*,x,y,z,J) \times \nonumber\\&&  \prod_{i=1}^{x}\frac{\Gamma _n(E_i^*,J)}{\Gamma _{\rm tot}(E_i^*,J)} \prod_{j=1}^{y}\frac{\Gamma _p(E_j^*,J)}{\Gamma_{\rm tot}(E_i^*,J)} \prod_{k=1}^{z}\frac{\Gamma _\alpha (E_k^*,J)}{\Gamma _{\rm tot}(E_k^*,J)}, 
\end{eqnarray}
where the $E_{\rm CN}^*$ and $J$ were the excitation energy and the spin of the excited nucleus, respectively. The total width $\Gamma_{\rm tot}$ was the sum of partial widths of particle evaporation, $\gamma$-rays, and fission. The excitation energy $E_S^*$ before evaporating the $s$-th particles was evaluated by
\begin{eqnarray}
E_{s+1}^*=E_s^* - B _i ^n - B _j ^p - B_k ^\alpha - 2T_s
\end{eqnarray}
with the initial condition $E\rm_i^*$=$E_{\rm CN}^*$ and $s$=$i$+$j$+$k$. The $B_{\rm i}^n$, $B_{\rm j}^p$, $B_{\rm k}^\alpha$ are the separation energy of the $i$-th neutron, $j$-th proton, $k$-th alpha, respectively. The nuclear temperature $T_i$ was defined by $E_{\rm i}^*=\alpha T_{\rm i}^2-T_{\rm i}$ with the level density parameter $a$. The decay width of the $\gamma$-rays and the particle decay were evaluated with a similar method in Ref. \cite{Chen_2016}.
The $P(E_{\rm CN}^*,J)$ is the realization probability of evaporation channels.  For one particle channel, the $P(E_{\rm CN}^*,J)$ is written as 
\begin{eqnarray}
P(E_{\rm CN}^*,J)= \exp\left ( -\frac{(E_{\rm CN}^*-B_s-2T)^2}{2\sigma ^2}  \right ) 
\end{eqnarray}
where $\sigma$ was taken as 2.5 MeV, which is the half-height width of the excitation function of the residual nuclei. For the multiple neutrons channels $(x>1)$, the $P(E_{\rm CN}^*,J)$ could be derived by the Jackson formula\cite{doi:10.1139/p56-087,Chen_2017}.

The $\Gamma _n(E_i^*,J)$, $\Gamma _p(E_i^*,J)$ and $\Gamma _{\alpha}(E_i^*,J)$ are the decay widths of particles n, p, $\alpha$, which are evaluated by the Weisskopf evaporation theory\cite{PhysRevC.68.014616}. The fission width $\Gamma_f(E^*,J)$ was calculated by the Bohr-Wheeler formula\cite{PhysRevC.80.057601,artza05,PhysRevC.76.044606}.

In our calculation, the fission barrier has a microscopic part and the macroscopic part which is written as
\begin{eqnarray}
B\rm_f(E^*,J)=B_f^{LD}+B_f^M(E^*=0,J)\rm exp(-E^*/E_D)
\end{eqnarray}
where the macroscopic part was derived from the liquid-drop model, as follows
\begin{eqnarray}
B^{\rm LD}\rm_f = \left \{\begin{array} {rl} 0.38(0.75 - x )E\rm_{s0} & , (1/3 < x < 2/3) \\ \\
0.83(1-x)^3 E\rm_{s0} & ,(2/3 < x < 1) \end{array} \right.
\end{eqnarray}
with 
\begin{eqnarray}
x=\frac{E\rm_{C0}}{2E\rm_{S0}}. 
\end{eqnarray}
Here, $E_{\rm c0}$ and $E_{\rm s0}$ were the surface energy and Coulomb energy of the spherical nuclear, respectively, which could be taken from the Myers-Swiatecki formula
\begin{eqnarray}
E_{\rm s0}=17.944[1-1.7826(\frac{N-Z}{A})^2]A^{2/3} \ MeV 
\end{eqnarray}
and
\begin{eqnarray}
E_{\rm c0}=0.7053\frac{Z^2}{A^{1/3}} \ MeV. 
\end{eqnarray}
Microcosmic shell correction energy was taken from \cite{Moller_1995}. Shell-damping energy was taken as $E_{\rm D}=5.48A^{1/3}/(1+1.3A^{-1/3})$ MeV.

\begin{table*}[!ht]
\tabcolsep=.06 cm
\renewcommand\arraystretch{1.2}
\centering
\footnotesize
\caption{\label{tab2} The possible combinations of projectiles-targets for synthesizing new superheavy nuclei. }
\begin{tabular}{c|ccccccccccc}
\hline
 T.$\backslash$ P.&$^{40-48}$Ca & $^{43-49}$Sc &$ ^{44-50}$Ti &  $^{47-51}$V & $ ^{50-54}$Cr & $^{51-56}$Mn &$^{53-60}$Fe &$^{56-61}$Co & $^{58-64}$Ni &$^{63-65}$Cu &$^{64-68}$Zn \\ \hline
$^{232}$Th &$ ^{272-280}$Ds &$^{275-281}$Rg &$ ^{276-282}$Cn &$ ^{279-283}$Nh & $^{282-286}$FI &$ ^{283-288}$Mc &$ ^{285-292}$Lv & $^{288-293}$Ts &$^{290-296}$Og & $^{295-297}$119&$^{296-300}$120   \\ 
$^{231}$Pa & $^{271-279}$Rg & $^{274-280}$Cn &$ ^{275-281}$Nh & $^{278-282}$FI & $^{281-285}$Mc & $^{282-287}$Lv & $^{284-291}$Ts &$^{287-291}$Og& $^{289-295}$119&$^{294-296}$120 & $^{295-299}$121 \\ 
$^{238}$U &$ ^{278-286}$Cn & $^{281-287}$Nh & $^{282-288}$FI & $^{285-289}$Mc &$ ^{288-292}$Lv &$ ^{289-294}$Ts &$^{291-298}$Og & $^{294-298}$119&$^{296-303}$120 & $^{301-303}$121 &$^{302-306}$122 \\ 
$^{237}$Np & $^{277-285}$Nh & $^{280-286}$FI & $^{281-287}$Mc & $^{284-288}$Lv &$ ^{287-291}$Ts &$^{288-293}$Og& $^{290-297}$119&$^{293-297}$120 & $^{295-301}$121 & $^{300-302}$122 & $^{301-305}$123 \\ 
$^{244}$Pu &$^{284-292}$FI & $^{287-293}$Mc & $^{288-294}$Lv & $^{291-295}$Ts &$^{294-298}$Og& $^{295-300}$119&$^{297-304}$120 & $^{300-304}$121 & $^{302-308}$122 & $^{307-309}$123 & $^{308-312}$124 \\ 
$^{243}$Am & $^{283-291}$Mc & $^{286-292}$Lv &$ ^{287-293}$Ts &$^{290-294}$Og& $^{293-297}$119&$^{294-299}$120 & $^{296-303}$121 & $^{299-303}$122 & $^{301-307}$123 & $^{306-308}$124 & $^{307-311}$125 \\ 
$^{247}$Cm & $^{287-295}$Lv &$ ^{290-296}$Ts  &$^{291-297}$Og& $^{294-298}$119&$^{297-301}$120 & $^{298-303}$121 & $^{300-307}$122 & $^{303-307}$123 & $^{305-311}$124 & $^{310-312}$125 & $^{311-315}$126 \\ 
$^{247}$Bk &$ ^{287-295}$Ts &$^{290-296}$Og& $^{291-297}$119&$^{294-298}$120 & $^{297-301}$121 & $^{298-303}$122 & $^{300-307}$123 & $^{303-307}$124 & $^{305-311}$125 & $^{310-312}$126 & $^{311-315}$127\\
$^{251}$Cf & $^{291-299}$Og& $^{294-300}$119&$^{295-301}$120 & $^{298-302}$121 & $^{301-305}$122 & $^{302-307}$123 &$ ^{304-311}$124 & $^{307-311}$125 & $^{309-315}$126 & $^{314-316}$127 &$^{315-319}$128 \\ 
$^{252}$Es & $^{292-300}$119&$^{295-301}$120 & $^{296-302}$121 & $^{299-303}$122 & $^{302-306}$123 & $^{303-308}$124 & $^{305-312}$125 & $^{308-312}$126 & $^{310-316}$127 &$^{315-317}$128 & $^{316-320}$129 \\ 
$^{257}$Fm & $^{297-305}$120 & $^{300-306}$121 & $^{301-307}$122 & $^{304-308}$123 & $^{307-311}$124 & $^{308-313}$125 & $^{310-317}$126 & $^{313-317}$127 &$^{315-321}$128 & $^{320-322}$129 & $^{321-325}$130 \\
\hline 
\end{tabular}
\end{table*}

\section{Results and discussion}
\label{sec3}
In previous work, the DNS model provides the calculations of the superheavy nuclei that have a good agreement with the available experimental data\cite{2011fennpr,FZQ2009,nst2021niu,FENG2010384c,PhysRevC.90.014612,PhysRevC.100.011601,PhysRevC.98.014618,PhysRevC.97.064609,PhysRevC.93.064610,PhysRevC.96.024610,PhysRevC.95.034323,PhysRevC.93.044615,PhysRevC.91.011603,PhysRevC.91.064612,PhysRevC.92.034612,PhysRevC.92.014601,doi:10.1142/S021830130801091X,PhysRevC.76.044606,FENG200650,PhysRevC.80.057601,PhysRevC.89.037601,PhysRevC.85.041601,Li_2006,PhysRevC.78.054607}, which has been used to predict production cross sections of the unknown SHEs located at the next periodic table of elements, based on the reactions of combinations of stable projectile-target. In this paper, we have performed systematic calculations of all possible combinations of available projectile-target particularly involving radioactive nuclei. The calculation results and details for the reactions of $^{48}$Ca+$^{257}$Fm, $^{56}$Mn+$^{243}$Am, $^{68}$Zn+$^{232}$Th, $^{48}$Ca+$^{252}$Es, $^{50}$Ti+$^{247}$Bk, $^{54}$Cr+$^{243}$Am, $^{50}$Ti+$^{251}$Cf, $^{54}$Cr+$^{247}$Cm are introduced as following graphs. All possible combinations of projectile-target have been demonstrated in Table \ref{tab2}, which have a long enough life to collide. The cross sections of 2n-, 3n-, 4n-, and 5n-evaporation channels for the reactions of the projectile from Ca to Zn bombarding on target from Ac to Fm have been exhibited in Table \ref{tab3}. From the calculation results, we found that the projectile beyond the Ni-induced reaction provides an extremely low cross section for the new superheavy elements with Z=119, 120. Therefore, the discussion has been made for the reactions of $^{48}$Ca+$^{252}$Es, $^{50}$Ti+$^{247}$Bk, $^{54}$Cr+$^{243}$Am and $^{48}$Ca+$^{257}$Fm, $^{50}$Ti+$^{251}$Cf, $^{54}$Cr+$^{247}$Cm at wide excitation energy.
\begin{figure*}[htbp!]
  \centering
  \includegraphics[width=.8\textwidth]{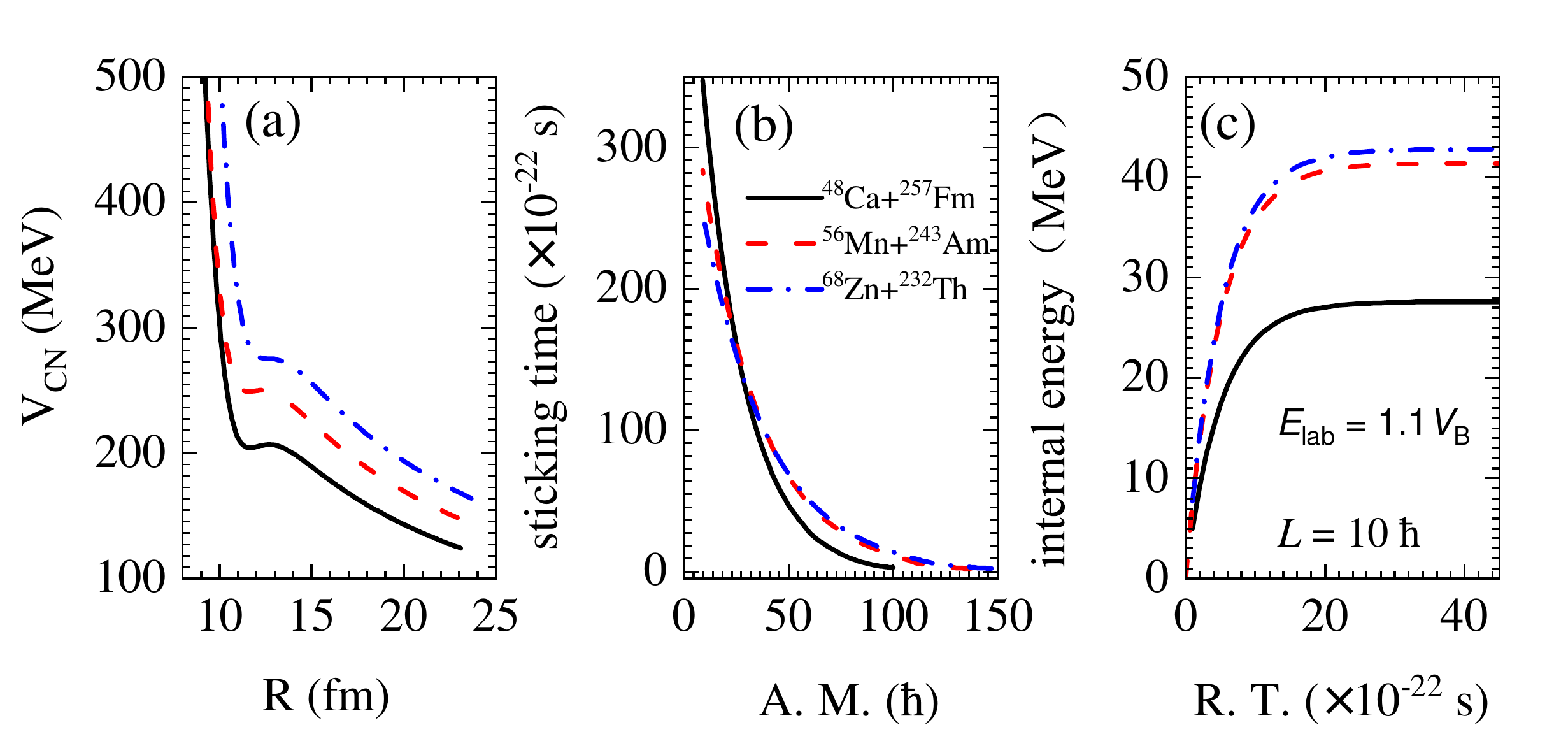}
  \caption{(Color online) The black solid lines, red dash lines, and blue dash-dot lines stand for the reactions $^{48}$Ca+$^{257}$Fm, $^{56}$Mn+$^{243}$Am, $^{68}$Zn+$^{232}$Th, respectively. Panel (a) shows interaction potential along with the distance. The panel (b) represents the sticking time to angular momentum (A. M.) at the incident energy $E_{\rm lab}$ = 1.1$V_{\rm B}$. Panel (c) displays the internal excitation energy along with time for the given A.M. 10 $\hbar$.}
  \label{FG1}
\end{figure*}

To investigate the effect of the entrance channel on the colliding process, the interaction potential along distance, the sticking time along with angular momentum, and internal excitation energies were listed in Fig. \ref{FG1}, where the black solid lines and dash red lines and dash-dot blue lines stand for the reactions $^{48}$Ca+$^{257}$Fm ($\eta$=0.68), $^{56}$Mn+$^{243}$Am ($\eta$=0.62), $^{68}$Zn+$^{232}$Th ($\eta$=0.54), respectively. The $\eta$ is the mass asymmetry with respect to the $\eta = (A_T-A_P)/(A_T+A_P)$ where the $A_T$ and $A_P$ are the mass of the target and projectile, respectively. Panel (a) shows the interaction potential energy of the different combinations that lead to the same element being so different, which because the collisions with the smaller $\eta$ lead to the larger Coulomb force. Panel (b) shows the sticking time varies along with the angular momentum, which has been calculated by the parametrization deflection function.
The parametrization deflection function is highly dependent on the interaction potential. It was found that the collisions with deep pocket potential have a long sticking time. For given impact parameter $L$ = 10 $\hbar$, internal excitation energies evolve with time, which accumulates from kinetic energy dissipation. The internal excitation energy reaches the equilibrium state at the time about 2$\times 10^{-22}$ s. At the equilibrium states, the collisions with large $\eta$ own the large internal excitation energy which is related to the interaction potential or the structure of colliding partners.  The interaction potential, and sticking time are the basic input physic quantities in the DNS model. The internal excitation energy could reveal the dynamic diffusion process.  

\begin{figure*}[htbp!]
  \centering
  \includegraphics[width=.8\textwidth]{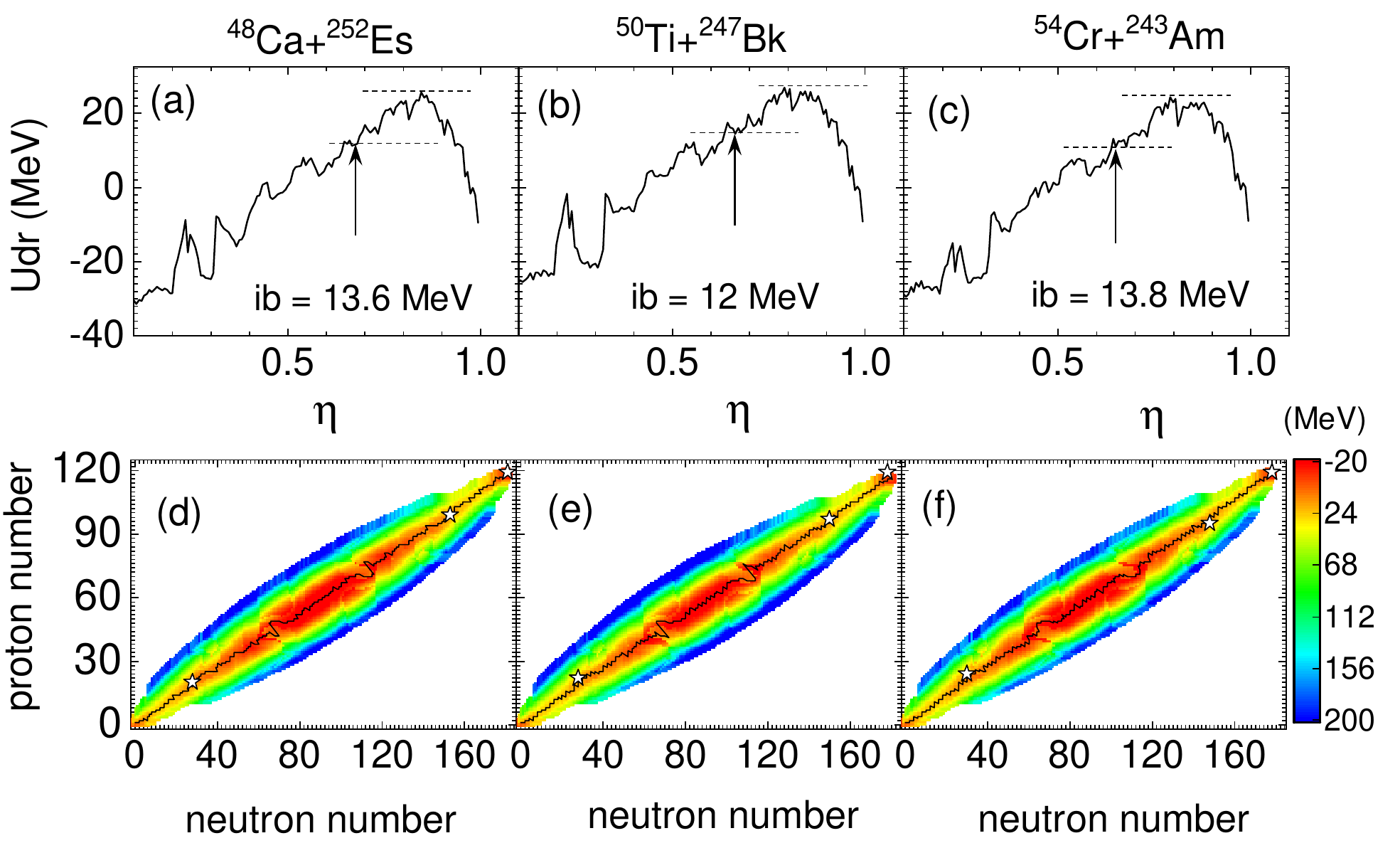}
  \caption{(Color online) The driving potential and potential energy surface of the reactions $^{48}$Ca+$^{252}$Es, $^{50}$Ti+$^{247}$Bk, $^{54}$Cr+$^{243}$Am are shown as the up-panels and bottom-panels, respectively.}
  \label{FG21}
\end{figure*}
In the DNS model, the colliding partners overcome the Coulomb barrier to form the composite system, where the relative kinetic energy and angular momentum dissipated into the internal excitation energy which enables the nucleons to transfer between the touching projectile and target. The probability of nucleon transfer is calculated by solving a set of master equations in which the potential energy surface (PES) or the driving potential are the crucial input physical quantities. The PES was derived by Eq. (\ref{pes}) which contained the ground state Q value, the interaction potential, and the rotational energy at the head-on collision with the fixed distance. The driving potential and PES of the reactions $^{48}$Ca+$^{252}$Es ($\eta$=0.68), $^{50}$Ti+$^{247}$Bk ($\eta$=0.66), $^{54}$Cr+$^{243}$Am ($\eta$=0.63) lead to the new element with atomic number Z=119 were listed in Fig. \ref{FG21} as the three lists of panels. The up-panels and bottom panels were the driving potential and PES, respectively. In the up-panels, the arrow lines and the bottom dash lines stand for the injection points. The up dash lines represented the Businaro-Gallone (B.G.) points. The barrier from the injection point to the B.G. point was named the inner fusion barrier. The inner barriers of the $^{48}$Ca+$^{252}$Es ($\eta$=0.68), $^{50}$Ti+$^{247}$Bk ($\eta$=0.66), $^{54}$Cr+$^{243}$Am ($\eta$=0.63) were 13.6 MeV, 12 MeV, 13.8 MeV. It was found that the reaction of $^{50}$Ti+$^{247}$Bk ($\eta$=0.66) owned the smaller inner fusion barrier which might result in the large fusion probability. The open stars stand for the injection points and compound nuclei in the bottom panels where the valley trajectories as the solid black lines were added. We could see that the injection points were located along with the driving potential which was close to the $\beta$-stable line. There was a big pocket in the symmetry region of the PES. The PESs might clearly show the paths of the colliding system diffusion. 

\begin{figure*}[htbp!]
  \centering
  \includegraphics[width=.8\textwidth]{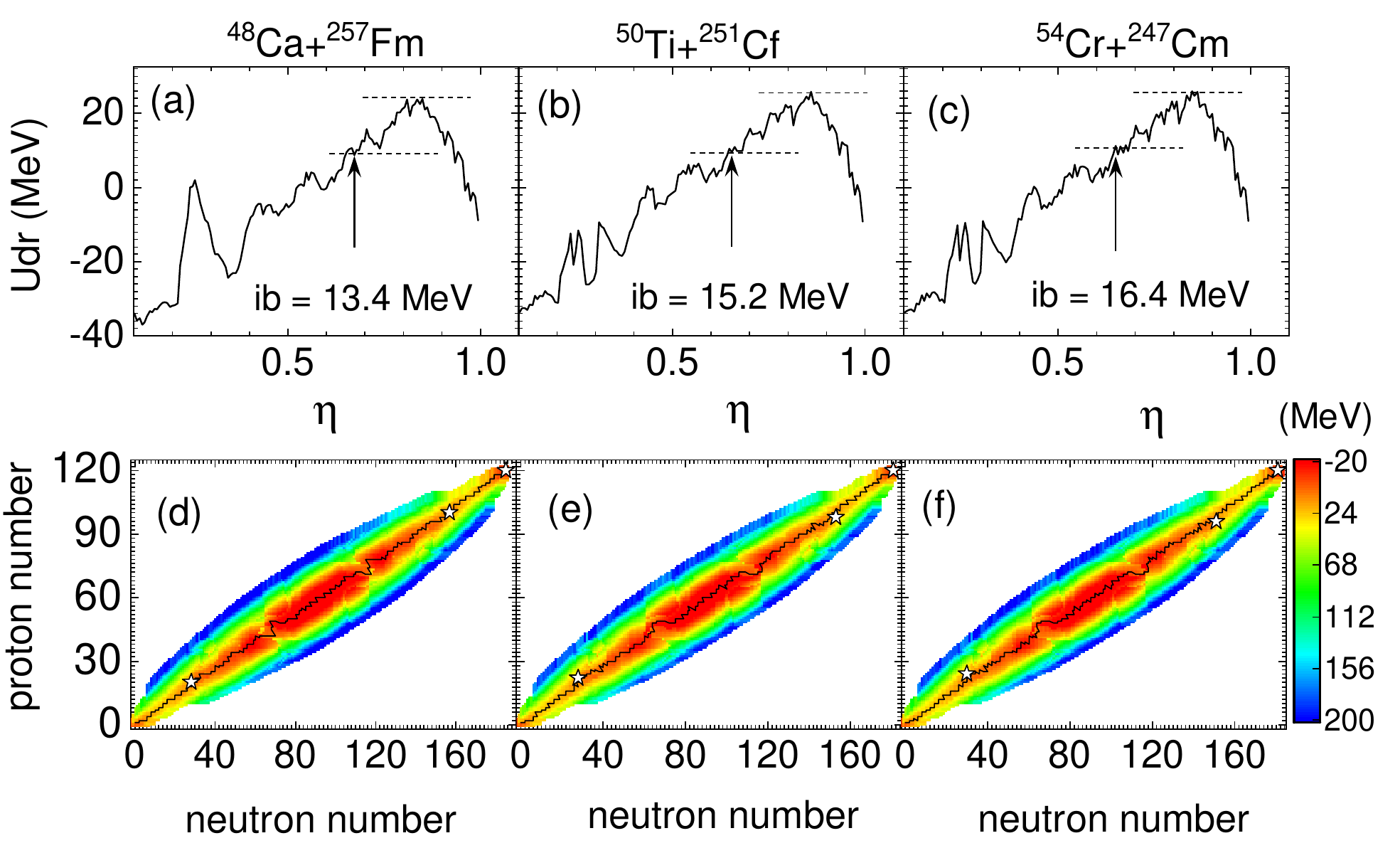}
  \caption{(Color online) The driving potential and potential energy surface (PES) of the reactions $^{48}$Ca+$^{257}$Fm, $^{50}$Ti+$^{251}$Cf, $^{54}$Cr+$^{247}$Cm are shown as the up-panels and bottom-panels, respectively.}
  \label{FG22}
\end{figure*}

Figure \ref{FG22} shows three lists of panels from left to right that were the driving potential and PES of the reactions $^{48}$Ca+$^{257}$Fm ($\eta$=0.68), $^{50}$Ti+$^{251}$Cf ($\eta$=0.66), $^{54}$Cr+$^{247}$Cm($\eta$=0.64) at the head-on collision orientation with the frozen distance, respectively, which has similar content to the Fig. \ref{FG21}. 
The up-panels presented the inner fusion barriers of $^{48}$Ca+$^{257}$Fm ($\eta$=0.68), $^{50}$Ti+$^{251}$Cf ($\eta$=0.66), $^{54}$Cr+$^{247}$Cm($\eta$=0.64) owned the values 13.4 MeV, 15.2 MeV, 16.4 MeV, respectively.
It was found that the inner fusion barrier were increased with the decreased mass asymmetry system, which could be used to evaluate the relative fusion probability. Compared to the colliding systems used to produce new element Z =119, the inner barriers belonging to the colliding system for synthesizing new element Z =120 have a larger value, which means that element Z = 120 has a lower production cross-section than that of element Z =119. The bottom panels (d), (e), and (f) exhibited the PESs which could be applied to predict the tendency of probability diffusion.

\begin{figure*}[htbp!]
  \centering
  \includegraphics[width=0.8\textwidth]{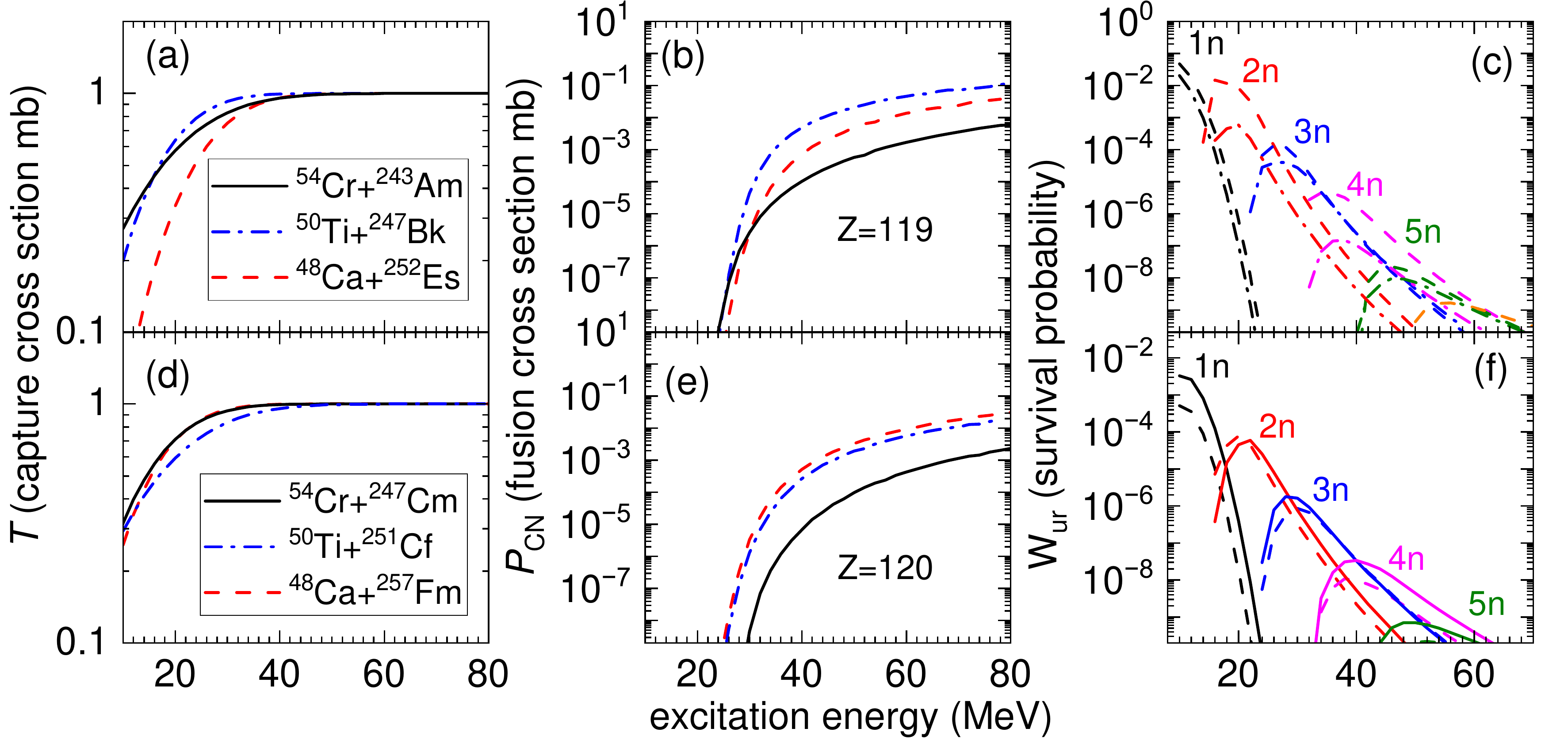}
  \caption{(Color online) For the reactions of $^{48}$Ca+$^{257}$Fm, $^{50}$Ti+$^{251}$Cf, $^{54}$Cr+$^{247}$Cm and and the reactions of $^{48}$Ca+$^{252}$Es, $^{50}$Ti+$^{247}$Bk, $^{54}$Cr+$^{243}$Am lead to new element with atomic number Z=119, 120 at wide incident energies, respectively. The capture cross-sections, fusion cross-section, and survival probability are listed in the three lists of panels.} 
  \label{FG3}
\end{figure*}

The DSN model classifies the fusion-evaporation reactions into three stages i.e. the capture stage, the fusion stage, and the survival stage, which was calculated based on one dimension barrier penetration model or the Hill-Wheeler formula (Eq. \ref{hwt}), master equation and evaporation statistics model, respectively, correspond to the three lists of panels from the left to right in Fig. \ref{FG3}, where the projectile $^{48}$Ca, $^{50}$Ti and $^{54}$Cr induced reactions were marked by red dash lines, blue dash-dot lines, and black solid lines, respectively. The up-panels were for the reactions of $^{48}$Ca+$^{252}$Es, $^{50}$Ti+$^{247}$Bk, $^{54}$Cr+$^{243}$Am lead to the new superheavy element Z=119. The capture cross-section of the head-on collisions of $^{48}$Ca+$^{252}$Es, $^{50}$Ti+$^{247}$Bk, $^{54}$Cr+$^{243}$Am were listed in panel (a). It was found that the penetration probability in descending order was $^{50}$Ti+$^{247}$Bk, $^{54}$Cr+$^{243}$Am, $^{48}$Ca+$^{252}$Es in the sub-barrier region, which did not respect to their mass asymmetry order. Because the deformation of colliding partners coupled with the Coulomb force plays a crucial role in the penetration process. Panel (b) shows the fusion probability from top to bottom where the reactions $^{50}$Ti+$^{247}$Bk, $^{48}$Ca+$^{252}$Es, $^{54}$Cr+$^{243}$Am, respectively, which has consistency with the inner fusion barrier order. In the framework of the DNS model, complete fusion reactions were treated as all the nucleons transferred from the projectile to the target. In the diffusion process, the formation probability of all possible fragments was evaluated by solving a set of master equations. The fragments reaching the B.G. point would be considered as overcoming the inner fusion barrier, which was deemed as the fusion event. The reactions $^{48}$Ca+$^{252}$Es, $^{54}$Cr+$^{243}$Am, $^{50}$Ti+$^{247}$Bk lead to the excited compound nuclei $^{300}$119, $^{297}$119, respectively, whose survival probability of 2n-, 3n-, 4n-, 5n-evaporation channels were listed in Fig. \ref{FG3}, marked by the black, red, blue, magenta, olive lines, respectively. We found that the survival probability of more neutron number channels decreased rapidly, due to the fission decay being dominant in this nuclei region. The capture cross-sections of the reactions  $^{48}$Ca+$^{257}$Fm ($\eta$=0.68), $^{50}$Ti+$^{251}$Cf ($\eta$=0.66), $^{54}$Cr+$^{247}$Cm($\eta$=0.64) for producing new element Z=120 were listed in panel (d), marked by red dash lines, blue dash-dot lines and solid black lines, respectively, which did not have a big difference. Because of the coupled effect of the strong Coulomb force and the deformation. Their compound nuclei are $^{305}$120, $^{301}$120. The fusion probability was listed in Fig. \ref{FG3}. 
The two highly excited compound nuclei de-excited by evaporating light particles. The evaporation probability of 1n-, 2n-, 3n-, 4n-, and 5n-evaporation channels for the $^{305}$120, $^{301}$120 were shown in panel (f).
It was found that the survival probability was relatively small, compared to the other two stages, which depress the synthesis cross sections strongly. Therefore, the predicted synthesis cross-sections of new SHE were highly dependent on the survival probability, where shell correction energy plays a crucial role. 
\begin{figure*}[htbp!]
  \centering
  \includegraphics[width=0.8\textwidth]{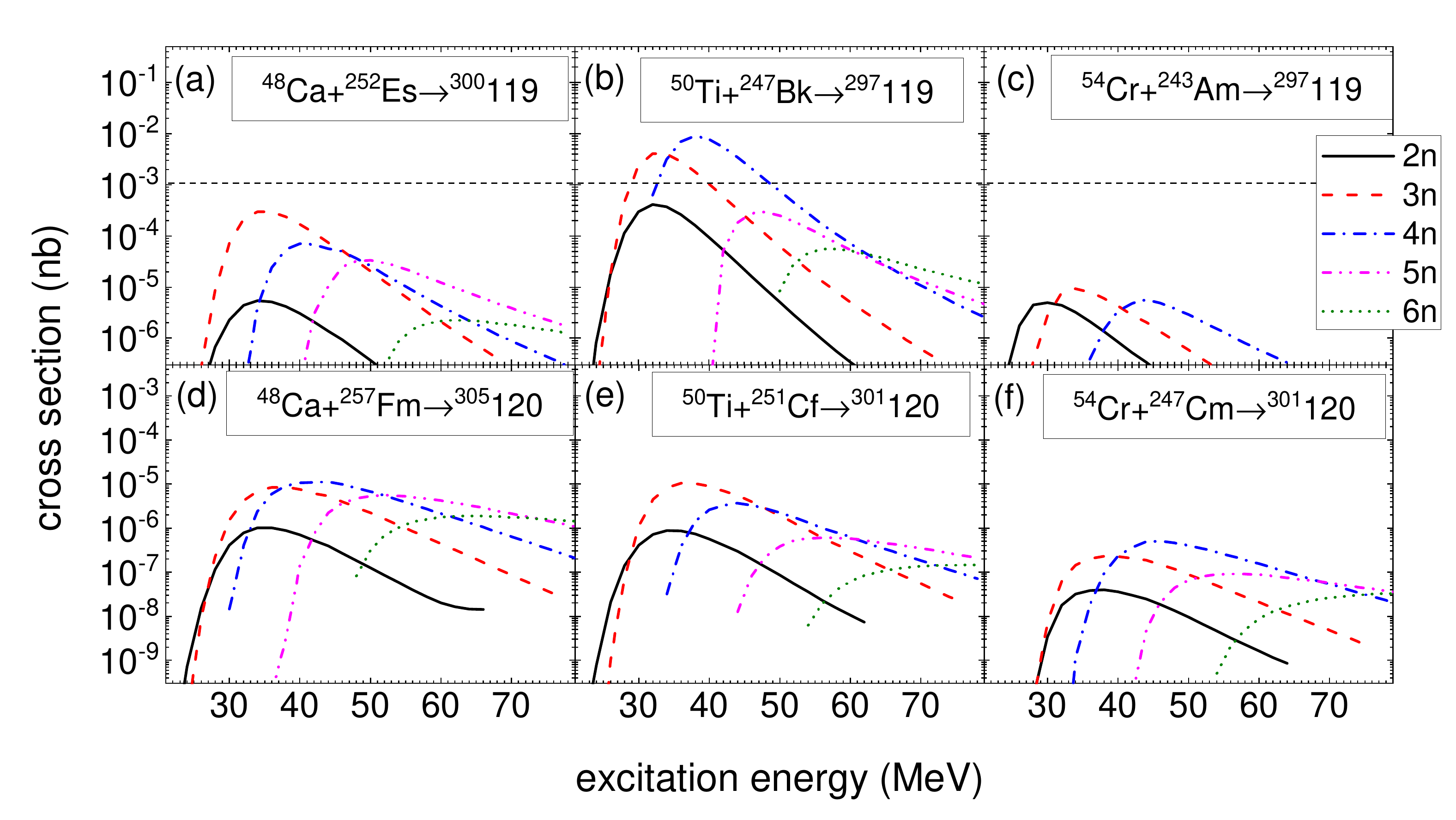}
  \caption{(Color online) The evaporation residual cross sections of $^{48}$Ca+$^{252}$Es$\rightarrow$$^{300}$119, $^{50}$Ti+$^{247}$Bk$\rightarrow$$^{297}$119, $^{54}$Cr+$^{243}$Am$\rightarrow$$^{297}$119 are listed in up-panels. The evaporation residual cross sections of $^{48}$Ca+$^{257}$Fm$\rightarrow$$^{305}$120, $^{50}$Ti+$^{251}$Cf$\rightarrow$$^{301}$120, $^{54}$Cr+$^{247}$Cm$\rightarrow$$^{301}$120 are listed in bottom-panels.}
  \label{FG4}
\end{figure*}

We multiplied the capture cross-section, the fusion probability, and the survival probability together to get the residual cross-sections of 2n-, 3n-, 4n-, 5n-, and 6n-evaporation channels which were marked by solid black lines, red dash lines, blue dash-dot lines, magenta dash-dot-dot lines, respectively, as shown in Fig. \ref{FG4}. The up-panels show the excitation functions of multiple neutron evaporation channels of the reactions $^{48}$Ca+$^{252}$Es$\rightarrow$$^{300}$119, $^{50}$Ti+$^{247}$Bk$\rightarrow$$^{297}$119, $^{54}$Cr+$^{243}$Am$\rightarrow$$^{297}$119. Panel (a) shows that the largest cross-section is 0.3 nb in the reactions $^{48}$Ca($^{252}$Es, 3n)$^{297}$119 at the excitation energy 34 MeV. Panel (b) shows that the largest cross-section is 8 nb in the reactions $^{50}$Ti($^{247}$Bk, 4n)$^{293}$119 at the excitation energy 38 MeV. Panel (c) shows that the largest cross-section is 0.01 nb in the reactions $^{54}$Cr($^{243}$Am, 3n)$^{294}$119 at the excitation energy 34 MeV. It was found that the reaction of $^{50}$Ti+$^{247}$Bk at excitation 38 MeV is the best to produce new element Z=119. The production cross-sections of the new superheavy element Z=119 are highly dependent on the entrance channels. 
The bottom panels show the excitation functions of multiple neutron evaporation channels of the reactions $^{48}$Ca+$^{257}$Fm$\rightarrow$$^{305}$120, $^{50}$Ti+$^{251}$Cf$\rightarrow$$^{301}$120, $^{54}$Cr+$^{247}$Cm$\rightarrow$$^{301}$120. Panel (d) displays the largest cross-section is 0.01 nb in the reactions $^{48}$Ca ($^{257}$Fm, 4n) $^{301}$120 at the excitation energy 42 MeV, where the neutron number of the synthesized superheavy isotope $^{301}$120 N=181 is so close to the superheavy island of stability. 
Panel (e) displays that the largest cross-section is 0.01 nb in the reactions $^{50}$Ti ($^{251}$Cf, 3n) $^{298}$120 at the excitation energy 36 MeV. Panel (f) displays the largest cross-section is $5\times 10^{-3}$ nb in the reactions $^{54}$Cr ($^{247}$Cm, 4n) $^{297}$120 at the excitation energy 46 MeV.
Fig. \ref{FG4} exhibits the influence of entrance channels on production cross-sections of new superheavy isotopes. The best collision combinations of synthesizing new superheavy elements are the reactions of projectile $^{50}$Ti induced reaction on the target $^{247}$Bk, $^{251}$Cf.

\begin{figure*}[htbp!]
  \centering
  \includegraphics[width=.8\textwidth]{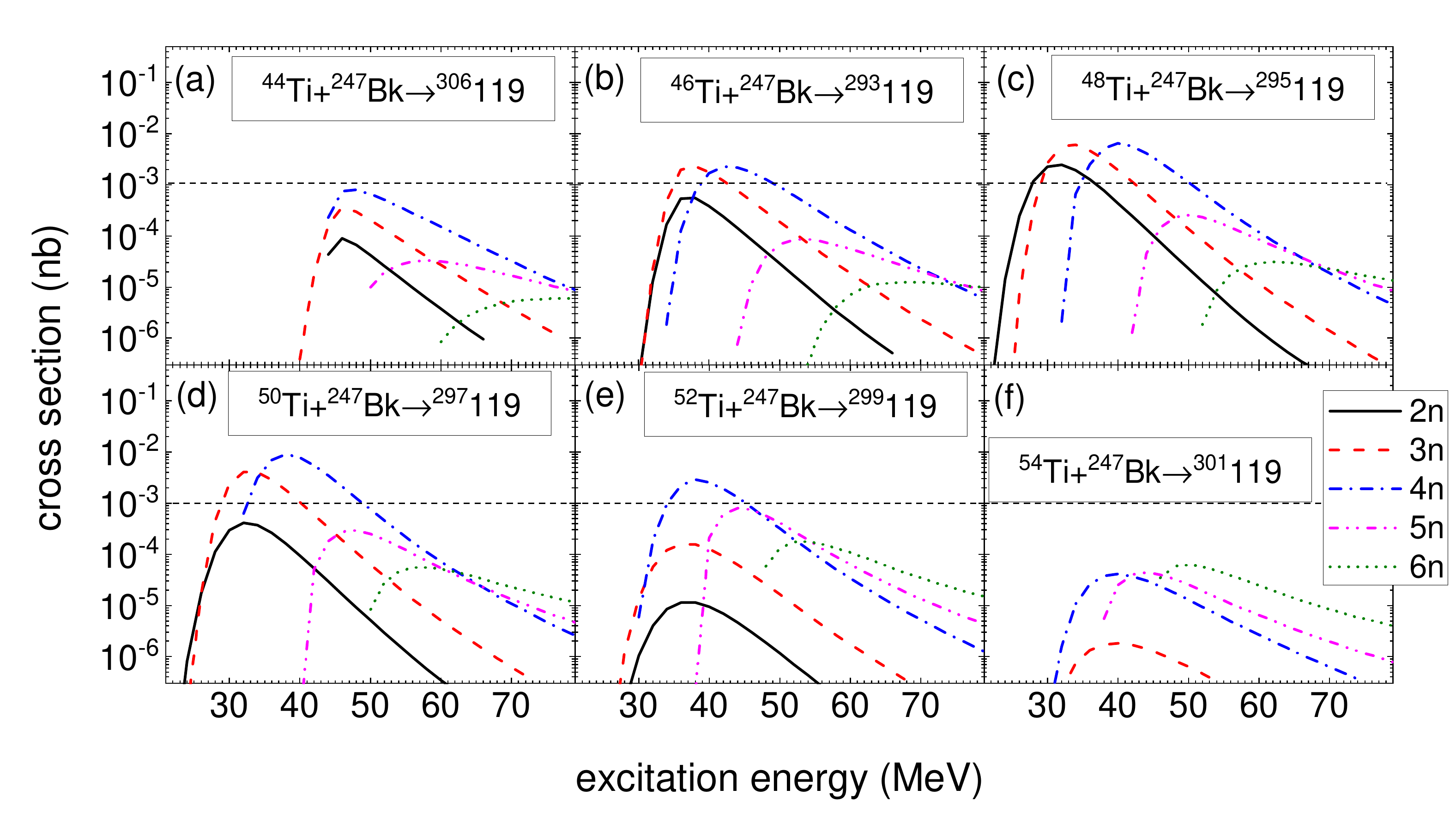}
  \caption{(Color online) The cross sections of 2n-, 3n-, 4n-, 5n-evaporation channels in the reactions of projectile $^{44,46,48,50,52,54}$Ti bombarding on target $^{247}$Bk are listed in these panels.}
  \label{FG5}
\end{figure*}

From the investigation of the entrance channel effect on the production cross-section of specific superheavy nuclei, the best projectile-target combination of synthesizing the element Z=119 was found $^{50}$Ti + $^{247}$Bk. The ratio of N/Z for projectile isotopes might contribute to the formation cross-section or depress the synthesis probability for the specific SHN. To investigate the dependence of residual cross-sections of multiple neutrons evaporation channels on the isospin of the projectile, the selections of $^{44,46,48,50,52,54}$Ti as the projectile bombarding on the target $^{247}$Bk have been made, which could lead to the new SHN with Z=119, as illustrated in Fig. \ref{FG5}, in which the black lines, red dash lines, blue dash-dot lines, magenta dash-dot-dot lines and short-dash lines indicate the excitation functions of the 2n-, 3n-, 4n-, 5n-,  and 6n-evaporation channels, respectively. Panel (a) shows the excitation function of the reactions $^{44}$Ti + $^{247}$Bk, in which the largest cross-section is 0.8 pb in the reactions $^{44}$Ti($^{247}$Bk, 4n)$^{293}$119 at the excitation energy 48 MeV. Panel (b) shows the largest cross-sections are the 2 pb, 2.1 pb, corresponding to the reactions $^{46}$Ti($^{247}$Bk, 3n)$^{290}$119 at the excitation energy 36 MeV, $^{46}$Ti($^{247}$Bk, 4n)$^{295}$119 at the excitation energy 48 MeV, respectively. Panel (c) shows the synthesis cross-sections of $^{293}$119, $^{292}$119, $^{291}$119 are 2 pb, 6 pb, 7 pb in the reactions $^{48}$Ti + $^{247}$Bk. Panel (d) presents the production cross sections of $^{294}$119, $^{293}$119 are 4 pb, and 8 pb, respectively. Panel (e) exhibits that the 4n- and 5n-evaporation channels are dominant in the reactions $^{52}$Ti + $^{247}$Bk, and have cross-sections of 3 pb, 0.8 pb. The $^{54}$ Ti-induced reactions produce very low cross sections of SHN ($\sigma <$ 0.1 pb). For producing new SHE with Z=119, the Ti-induced reactions present a high dependence of the production cross sections on the isospin of the projectile. We found that the collisions of $^{48}$Ti + $^{247}$Bk and $^{50}$Ti + $^{247}$Bk are suitable to synthesize new SHE with Z=119. 
\begin{figure*}[htbp!]
  \centering
  \includegraphics[width=.8\textwidth]{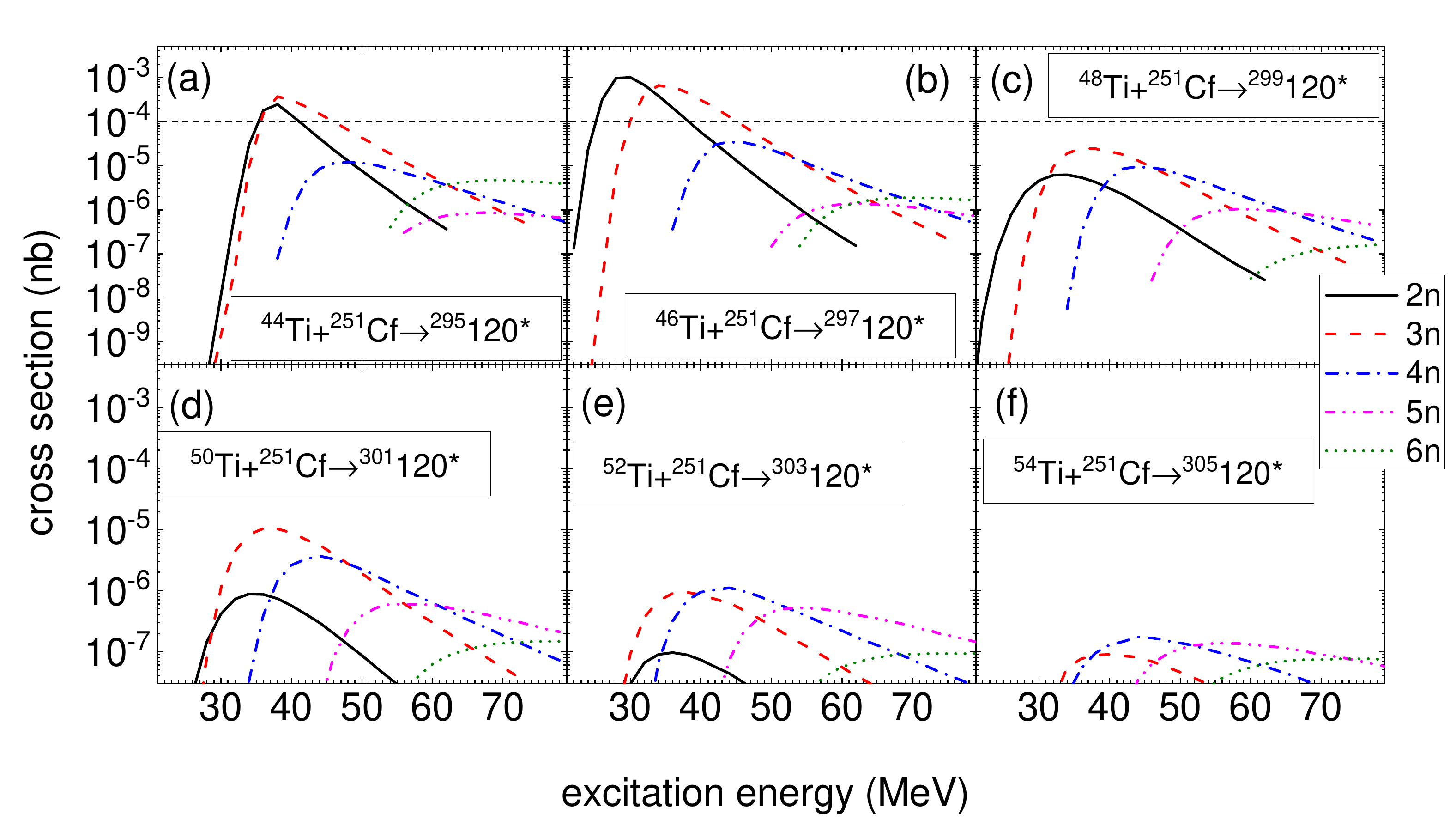}
  \caption{(Color online) The cross sections of 2n-, 3n-, 4n-, 5n-evaporation channels in the reactions of projectile $^{44,46,48,50,52,54}$Ti bombarding on target $^{251}$Cf are listed in these panels.}
  \label{FG6}
\end{figure*}

To find the best combinations of the projectile Ti isotopes and target $^{251}$Cf for synthesizing new SHE with Z=120, we have performed the calculations of the Ti-induced reactions systematically, as shown in Fig. \ref{FG6}, where the black lines, red dash lines, blue dash-dot lines, magenta dash-dot-dot lines and short-dash lines indicate the excitation functions of the 2n-, 3n-, 4n-, 5n-,  and 6n-evaporation channels, respectively. The black dash lines mark the cross-section 0.1 pb in the up-panels.  Fig. \ref{FG6} has a similar form to Fig. \ref{FG5}. Panel (a) is for the reactions $^{44}$Ti + $^{251}$Cf, which shows the largest cross section is around 0.2 pb for the 3n-channel. Panel (b) displays the $^{46}$Ti-induced reactions, where the 2n- and 3n-channels are dominant. The largest cross-section of the 2n-, and 3n-evaporation channels are about 1 pb, 0.6 pb in the reactions $^{46}$Ti ($^{251}$Cf, 2n) $^{295}$120 and $^{46}$Ti ($^{251}$Cf, 3n) $^{294}$120, respectively. The projectile $^{48}$Ti-induced reactions could lead to the synthesis cross-section 0.02 pb for 3n-channel, as shown in panel (c). Panels (d), (e), (f) show the $^{50,52,54}$ Ti-induced reactions that produced the extremely low cross sections ($<0.01$ pb). 
Figure \ref{FG6} has shown that production cross sections of new SHE with Z=120 are highly dependent on the isospin of the projectile. The most suitable combination of projectile-target is the reaction $^{46}$Ti + $^{251}$Cf which could synthesize SHN of $^{293}$120 and $^{294}$120 with the cross sections of 1 pb, 0.6 pb, respectively. 
\begin{table}[!ht]
\tabcolsep=.06 cm
\renewcommand\arraystretch{1.}
\centering
\footnotesize
\caption{\label{tab3} The production mechanisms of new superheavy nuclei with Z=119, 120 are listed as reactions, evaporation channels, excitation energy, and cross sections. }
\begin{tabular}{cccc|cccc}
\hline
 reaction & E.C. & $E^*$& $\sigma$ (pb) & reaction & E.C. & $E^*$ &$\sigma$ (pb) \\ \hline
 $^{48}$ Ca+$^{252}$Es & 2n & 34 & 5.4$\times$10$^{-3}$ &  $^{48}$ Ca+$^{257}$Fm  & 2n & 36 & 1$\times$10$^{-3}$ \\
   & 3n & 34 & 0.3 &    & 3n & 38 & 8.4$\times$10$^{-3}$ \\
   & 4n & 40 & 0.07 &    & 4n & 42 & 0.01 \\
   & 5n & 50 & 0.03 &    & 5n & 52 & 5.5$\times$10$^{-3}$ \\ \hline
 $^{49}$ Sc+$^{251}$Cf & 2n & 34 & 0.01 &  $^{49}$ Sc+$^{252}$Es  & 2n & 36 & 9$\times$10$^{-4}$ \\
   & 3n & 34 & 0.5 &    & 3n & 38 & 0.01 \\
   & 4n & 40 & 0.1 &    & 4n & 44 & 4.4$\times$10$^{-3}$ \\
   & 5n & 50 & 0.04 &    & 5n & 56 & 7.4$\times$10$^{-4}$ \\ \hline
 $^{50}$ Ti+$^{247}$Bk & 2n & 32 & 0.4 &  $^{50}$ Ti+$^{251}$Cf  & 2n & 34 & 8.7$\times$10$^{-4}$ \\
   & 3n & 32 & 4 &    & 3n & 36 & 0.01 \\
   & 4n & 38 & 8 &    & 4n & 44 & 3.7$\times$10$^{-3}$ \\
   & 5n & 48 & 0.3 &    & 5n & 56 & 6$\times$10$^{-4}$ \\ \hline
 $^{51}$ V+$^{247}$Cm & 2n & 34 & 0.04 &  $^{51}$ V+$^{247}$Bk  & 2n & 34 & 1.8$\times$10$^{-3}$ \\
   & 3n & 36 & 0.08 &    & 3n & 38 & 5.4$\times$10$^{-3}$ \\
   & 4n & 42 & 0.1 &    & 4n & 48 & 2.2$\times$10$^{-3}$ \\
   & 5n & 50 & 0.01 &    & 5n & 62 & 2.5$\times$10$^{-4}$ \\ \hline
 $^{54}$ Cr+$^{243}$Am & 2n & 30 & 5$\times$10$^{-3}$ &  $^{54}$ Cr+$^{247}$Cm  & 2n & 38 & 4$\times$10$^{-4}$ \\
   & 3n & 34 & 0.01 &    & 3n & 38 & 2.8$\times$10$^{-4}$ \\
   & 4n & 44 & 6$\times$10$^{-3}$ &    & 4n & 46 & 5$\times$10$^{-3}$ \\
   & 5n & 56 & 2.5$\times$10$^{-4}$ &    & 5n & 56 & 9$\times$10$^{-6}$ \\ \hline
 $^{56}$ Mn+$^{244}$Pu & 2n & 32 & 6$\times$10$^{-4}$ &  $^{56}$ Mn+$^{243}$Am  & 2n & 36 & 3.6$\times$10$^{-5}$ \\
   & 3n & 34 & 0.03 &    & 3n & 38 & 2$\times$10$^{-4}$ \\
   & 4n & 40 & 5.3$\times$10$^{-3}$ &    & 4n & 46 & 1.2$\times$10$^{-4}$ \\
   & 5n & 50 & 2.6$\times$10$^{-3}$ &    & 5n & 58 & 2.7$\times$10$^{-5}$ \\ \hline
 $^{60}$ Fe+$^{237}$Np & 2n & 32 & 1.4$\times$10$^{-3}$ &  $^{60}$ Fe+$^{244}$Pu  & 2n & 36 & 8$\times$10$^{-6}$ \\
   & 3n & 34 & 3$\times$10$^{-3}$ &    & 3n & 40 & 5.4$\times$10$^{-5}$ \\
   & 4n & 44 & 2.1$\times$10$^{-3}$ &    & 4n & 48 & 3$\times$10$^{-5}$ \\
   & 5n & 56 & 1$\times$10$^{-4}$ &    & 5n & 50 & 5$\times$10$^{-5}$ \\ \hline
 $^{61}$ Co+$^{238}$U & 2n & 34 & 2.7$\times$10$^{-4}$ &  $^{61}$ Co+$^{237}$Np  & 2n & 36 & 1.5$\times$10$^{-5}$ \\
   & 3n & 36 & 6.6$\times$10$^{-4}$ &    & 3n & 40 & 4.4$\times$10$^{-5}$ \\
   & 4n & 46 & 1.6$\times$10$^{-3}$ &    & 4n & 50 & 4.5$\times$10$^{-5}$ \\
   & 5n & 52 & 2.8$\times$10$^{-4}$ &    & 5n & 60 & 1$\times$10$^{-5}$ \\ \hline
 $^{64}$ Ni+$^{231}$Pa & 2n & 36 & < 10$^{-6}$ &  $^{64}$ Ni+$^{238}$U  & 2n & 36 & < 10$^{-6}$ \\
   & 3n & 38 & < 10$^{-6}$&    & 3n & 38 & < 10$^{-6}$ \\
   & 4n & 42 & < 10$^{-6}$ &    & 4n & 42 & < 10$^{-6}$ \\
   & 5n & 50 & < 10$^{-6}$ &    & 5n & 50 & < 10$^{-6}$ \\ \hline
 $^{65}$ Cu+$^{232}$Th & 2n & 36 & 0.08 &  $^{65}$ Cu+$^{231}$Pa  & 2n & 20 & < 10$^{-6}$ \\
   & 3n & 38 & < 10$^{-6}$ &    & 3n & 20 & < 10$^{-6}$ \\
   & 4n & 42 & < 10$^{-6}$ &    & 4n & 20 & < 10$^{-6}$ \\
   & 5n & 50 & < 10$^{-6}$ &    & 5n & 20 & < 10$^{-6}$ \\ \hline
 $^{68}$ Zn+$^{232}$Ac & 2n & 20 & < 10$^{-6}$ &  $^{68}$ Zn+$^{232}$Th  & 2n & 20 & < 10$^{-6}$ \\
   & 3n & 38 & < 10$^{-6}$ &    & 3n & 20 & < 10$^{-6}$ \\
   & 4n & 42 & < 10$^{-6}$ &    & 4n & 20 & < 10$^{-6}$ \\
   & 5n & 50 & < 10$^{-6}$ &    & 5n & 20 & < 10$^{-6}$ \\ \hline
\hline 
\end{tabular}
\end{table}
\section{Conclusions}
\label{sec4}

Summarizing, we pick up all available nuclei as the possible combinations of projectile-target from the nuclei chart, as illustrated in Table \ref{tab3}. We have calculated the fusion-evaporation reactions of the projectile Ca, Sc, Ti, V, Cr Mn, Fe, Co, Ni Cu, Zn bombarding on the target $^{257}$Fm, $^{252}$Es, $^{251}$Cf, $^{247}$Bk, $^{247}$Cm, $^{243}$Am, $^{244}$Pu, $^{237}$Np, $^{238}$U, $^{231}$Pa, $^{232}$Th, $^{232}$Ac within the DNS model systematically. The predictions of cross sections for new SHE with Z=119, 120 were listed in Table \ref{tab3}. It was found that the mass of the projectile beyond $^{61}$Co leads to an extremely low cross-section. The interaction potential along distance, the sticking time along with angular momentum, and the evolution of internal excitation energy are exported from our calculations which are the details of our model.
The driving potential and PESs of the reactions leading to Z=119 and 120 could be used to predict the tendency of the probability diffusion. Three stages of the fusion-evaporation reactions have been discussed, respectively. We found that the fusion stage is crucial to evaluate the residual cross-sections of specific SHN. Production cross sections of new SHN with Z=119-120 are highly dependent on the entrance channels of the colliding system. The dependence of cross sections excitation functions of multiple neutron channels on the isospin of projectile Ti-induced reactions. In our calculations, the most suitable combinations of projectile-target for synthesizing new SHE with atomic number Z=119,120 are the $^{46,48,50,52}$Ti + $^{247}$Bk and $^{46}$Ti+$^{251}$Cf. The new elements of Z=119, 120 have been predicted with synthesis cross sections, such as $^{289}$119 with $\sigma_{\rm 4n}$ = 2.3 pb in $^{46}$Ti ($^{247}$Bk, 4n,) $^{289}$119 at $E^*$ = 42 MeV, $^{290}$119 with $\sigma_{\rm 3n}$ = 2.2 pb in $^{46}$Ti ($^{247}$Bk, 3n,) $^{290}$119 at $E^*$ = 38 MeV, $^{291}$119 with $\sigma_{\rm 4n}$ = 2.5 pb in $^{48}$Ti ($^{247}$Bk, 4n,) $^{291}$119 at $E^*$ = 32 MeV, $^{292}$119 with $\sigma_{\rm 3n}$ = 6 pb in $^{48}$Ti ($^{247}$Bk, 3n,) $^{292}$119 at $E^*$ = 34 MeV, $^{293}$119 with $\sigma_{\rm 2n}$ = 6.5 pb in $^{48}$Ti ($^{247}$Bk, 2n,) $^{293}$119 at $E^*$ = 40 MeV, $^{294}$119 with $\sigma_{\rm 3n}$ = 4.1 pb in $^{50}$Ti ($^{247}$Bk, 3n,) $^{294}$119 at $E^*$ = 32 MeV, $^{295}$119 with $\sigma_{\rm 4n}$ = 2.9 pb in $^{52}$Ti ($^{247}$Bk, 4n,) $^{295}$119 at $E^*$ = 38 MeV, and $^{294}$120 with $\sigma_{\rm 3n}$ = 0.6 pb in $^{46}$Ti ($^{251}$Cf, 3n,) $^{294}$120 at $E^*$ = 34 MeV, $^{295}$120 with $\sigma_{\rm 2n}$ = 1 pb in $^{46}$Ti ($^{251}$Cf, 2n,) $^{295}$120 at $E^*$ = 30 MeV.
The optimal projectile-target combinations of $^{46}$Ti+$^{251}$Cf, $^{48}$Ti+$^{247}$Bk and beam energies of $E^*$ = 30-40 MeV for synthesizing new SHE with atomic number Z=119, 120 are proposed for the forthcoming experiments.
\section{Acknowledgments}
We acknowledge fruitful discussions with Prof. Zai-Guo Gan and Prof. Ming-Hui Huang for the motivation of this work. This work was supported by the National Natural Science Foundation of China ( 12105241, 12175072), the Natural Science Foundation of Jiangsu Province (No. BK20210788), the Jiangsu Provincial Double-Innovation Doctor Program (JSSCBS20211013) and the University Science Research Project of Jiangsu Province (21KJB140026) and Lv Yang Jin Feng (YZLYJFJH2021YXBS130) and the Key Laboratory of High Precision Nuclear Spectroscopy, Institute of Modern Physics, Chinese Academy of Sciences (No. IMPKFKT2021001).
%

\end{document}